\documentclass[conference]{IEEEtran}
\IEEEoverridecommandlockouts

\usepackage[normalem]{ulem}
\usepackage[ragged2e]{}
\usepackage{pgf-pie}
\usepackage{tikz}
\usepackage{enumerate}
\usepackage{multirow}
\usepackage{array}
\usepackage{listings}
\usepackage{subcaption}
\usepackage[hyphens]{url}
\usepackage{hyperref}
\usepackage[ruled,vlined]{algorithm2e}
\usepackage{amsmath}
\usepackage{amsfonts}
\usepackage{float}
\usepackage{pifont}
\usepackage[utf8]{inputenc}

\usepackage{upgreek}

\newcommand{\MYcomment}[1]{}
\renewcommand{\figurename}{Fig.}

\newcommand{\MYnote}[1]{}

\newcounter{MYtablecntr}
\addtocounter{MYtablecntr}{1}

\newcounter{MYalgorithmcntr}
\addtocounter{MYalgorithmcntr}{1}

\newcommand{\MYlabel}{\small {$\bullet$}}

\newcounter{MYenumctrtwo}

\newcounter{MYenumctr}

\newcommand{\code}[1]{\texttt{\small{#1}}}

\newcommand{\norm}[1]{\left\lVert#1\right\rVert}

\newcommand{\xmark}{\ding{55}}%

\newcommand{\px}[1]{{\href{https://github.com/PX4/Firmware/issues/#1}{\code{PX4-{#1}}}}}

\begin{document}
\raggedbottom

\title{Avis: In-Situ Model Checking for\\ Unmanned Aerial Vehicles}
\author{\IEEEauthorblockN{Max Taylor, Haicheng Chen, Feng Qin, Christopher Stewart}
\textit{The Ohio State University}}
\pagestyle{plain}

\maketitle

\begin{abstract}
Control firmware in unmanned aerial vehicles (UAVs) 
uses sensors to model and manage flight operations, from takeoff to landing 
to flying between waypoints.
However, sensors can fail at any time during a flight. 
If control firmware mishandles sensor failures, UAVs can 
crash,  fly away, or suffer other unsafe conditions.
In-situ model checking finds sensor failures that
could lead to unsafe conditions by systematically
failing sensors.  However, the type of 
sensor failure and its timing within a flight
affect its manifestation, creating a large search space.
We propose Avis, an in-situ model checker to 
quickly uncover UAV sensor failures that lead to unsafe conditions.
Avis exploits operating modes, i.e., a label that maps 
software execution to corresponding flight operations.   
Widely used control firmware already support operating modes.
Avis injects sensor failures as the control
firmware transitions between modes -- a key execution point
where mishandled software exceptions can trigger 
unsafe conditions.  
We implemented Avis and applied it to ArduPilot
and  PX4.  Avis found unsafe conditions 2.4X
faster than Bayesian Fault Injection, the leading,
state-of-the-art approach.  
Within the current code base of ArduPilot and PX4, Avis discovered 10 previously unknown software bugs that lead to unsafe conditions. Additionally, we reinserted 5 known bugs that caused serious, unsafe conditions and Avis correctly reported all of them.

\end{abstract}

\section{Introduction}
\label{sec:introduction}

Unmanned aerial vehicles (UAVs) hover, fly to set
waypoints and perform complex aerial operations.
Without a human aboard, UAVs can handle missions
that are too dangerous, too long or otherwise
unprofitable for traditional aircraft.
For example, UAVs can enter wildfires and war
zones~\cite{ambrosia2003demonstrating,ma2013simulation}.
UAVs can also survey large crop fields at low 
altitudes to assess damage caused by natural 
disasters, pests and contagious crop diseases~\cite{yang2020adaptive,zhang2020whole}.
UAVs use software, called
{\em control firmware}, to read from sensors, model the state of the aircraft,
respond to pilot commands,
and control pitch, thrust and yaw for navigation.
As the global market for UAVs will soon exceed
\$42B~\cite{droneii}, control firmware is increasingly crucial
system software.  It underlies every major UAV
use case and must support a growing number of
flight operations.
Software bugs in UAV control firmware can have serious 
consequences, such as crashes.

\begin{figure}
    \centering
    \includegraphics{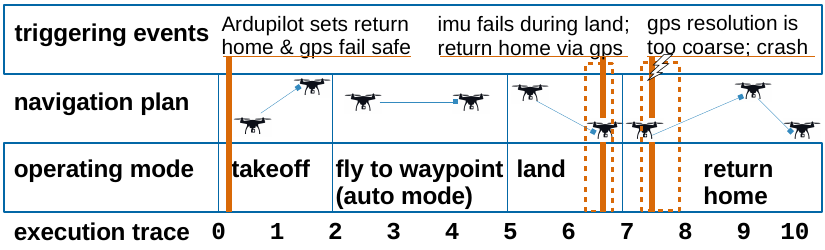}
    \caption{Execution analysis of a mishandled sensor failure that can cause a crash.}
    \label{fig:draw-bugexample}
\end{figure}

UAVs use a myriad of sensors, including 
inertial measurement units (IMUs),
global positioning systems (GPSs), 
compasses, and barometers.  Sensors
can fail for many reasons.  GPSs can be 
disrupted by Carrington events that 
affect the Earth's magnetic field~\cite{selwood2020what}.
Alternatively, sensors can be disconnected
from their power source due to turbulence or motor vibrations.
Control firmware, by design, accounts for
sensor failures via (1) failing over to 
redundant sensors, (2) discarding invalid
readings and (3) employing multiple methods
with diverse input needs to model the state of 
the aircraft.  Despite these precautions, 
sensor failures can cause {\em unsafe flight conditions} 
where control firmware executes flight operations
that crash the UAV or disrupt its mission.
{ Unsafe flight conditions} occur when 
fault handling logic does not anticipate the 
context surrounding a sensor failure.


Figure~\ref{fig:draw-bugexample}
depicts an unsafe condition that stemmed 
from a sensor failure in ArduPilot.  At the end of a landing
operation, IMU sensors failed.  The control
firmware detected the failure and began to
return to home using GPS driven flight.
At normal flight altitudes, these actions 
are safe; the GPS would measure altitude less precisely 
than an IMU but sufficiently to conduct simple maneuvers when used with other models.
However, the control firmware triggered GPS-driven
flight under the incorrect 
assumption
that the UAV
could safely navigate to a new altitude using the GPS alone.  At low altitudes, such
as the end of a landing operation,
GPS is too imprecise to guide major altitude adjustments.
We can repeatedly trigger this crash in simulation
by failing IMU sensors when the UAV is fewer than
2 meters above ground.

{\em Sensor bugs} are segments of control firmware
source code that, if fixed, could eliminate an
unsafe flight condition caused by a sensor failure.  
Figure~\ref{fig:draw-bugexample} was caused by
a sensor bug.  If control firmware checked 
altitude before switching to GPS flight, the 
crash could have been avoided.  Instead, the
landing routine could have been allowed to 
complete normally.  

Our analysis of public Github repositories reveals
that sensor bugs represent 40\% of source code 
patches intended to fix UAV crashes.
Further, sensor bugs often lead to crashes
or other serious consequences.
However, the source code for control 
firmware is large and complex.  
In practice, software developers wait for 
users to report sensor bug manifestations 
before trying to understand root causes.  The severity
of sensor bug manifestations necessitates 
more preemptive approaches.
In-situ model checking systematically injects 
faults during simulated executions, searching  
for faults that cause the system to violate 
invariant properties.  While in-situ model 
checking enables preemptive analysis,  UAVs 
present unique challenges for its application.
First, control firmware accesses sensor readings 
frequently ($10^3$--$10^4$ times per second). Also, one or multiple types of sensors can fail at any moment.
This failure space is immense.
Second, sensor bug manifestations depend 
on the timing and type of failure. 
Figure~\ref{fig:draw-bugexample} depicts the
narrow window where an IMU failure can cause
a UAV to crash.  Practical in-situ model 
checking approaches must balance these conflicting
concerns. While statistics-driven fault
injection \emph{seems} necessary given the magnitude
of the search space, the sampling approaches
could miss fault injections that trigger time-sensitive bugs.

This paper presents Avis, an
\underline{a}erial-\underline{v}ehicle 
\underline{i}n-\underline{s}itu model checker.  
Avis exploits a common sentiment among control
firmware developers: Sensor bugs often stem from
failure handling logic that is too narrowly
tailored to specific operating modes.
These bugs are hard to detect because failure handling logic is implemented in different locations in the firmware ~\cite{developerref}.
Avis uses custom in-situ workloads that exercise
transitions between  operating modes and carefully
injects failures, across all types of sensors,
near the transitions between operating modes.
By exploiting operating modes, Avis finds a nice
balance.  It prioritizes injection sites likely to reveal bugs, but
also captures time-sensitive issues at the critical
boundaries between operating modes.
Compared to Bayesian Fault Injection (BFI) \cite{bfi},
a statistically guided model checker for autonomous
vehicles, Avis does not rely on statistical inference.  
BFI is more likely to trigger
unsafe conditions that occur in the main flight
mode, especially if unsafe conditions have occurred
in the past.  In contrast, Avis does not require
training data and can comprehensively explore
fault handling logic that spans operating modes.

We implemented Avis and applied it to two
open-source control firmware: ArduPilot and PX4. 
We compared it to BFI in terms of efficiency
(unsafe conditions found per simulation) and
efficacy (bugs uncovered).  
Avis found unsafe conditions 2.4X more efficiently
than BFI.  When we re-inserted 5 previously known
software bugs that caused serious, unsafe
conditions, Avis found unsafe conditions caused by
each bug.  BFI did not find any.  When we studied 
unsafe conditions that Avis found in the current
code base, we found 10 previously unknown software
bugs related to IMU and GPS failures (2 of which
have been confirmed by developers).

To summarize, our contributions are:
\begin{itemize}
    \item A study characterizing the frequency and 
    impact of sensor bugs in widely used open-source control firmware.
    \item A fault injection approach that exploits 
    operating modes in UAV for stratified breadth-first search.
    \item A framework for building UAV workloads
      that exercise operating modes.
    \item A prototype of Avis and experimental results on ArduPilot and PX4 that 
    reveal the efficiency and efficacy of our approach by 
    capturing previously known sensor bugs and uncovering 
    new, previously unknown bugs.
\end{itemize}

\MYnote{
Specifically, we conduct a fault-injection
(FI) campaign at each sensor reading site under a
given UAV workload. This approach has the benefit
of completeness with respect to the target
workload, as well as soundness. In contrast,
recent work relies on model-extraction techniques
to identify lucrative FI sites. These techniques
require an expensive extraction process and
renders the search space incomplete.

We address the first two challenges by using
stratified sampling to guide fault
injection. Specifically, we order the simulation
of fault injection scenarios based on the
vehicle's current operating mode. This allows us
to explore many portions of a scene to quickly
identify unsafe scenarios. The last challenge we
address by providing a programmable framework UAV
developers can use to specify safety invariants
and build portable workloads. 

}

\begin{figure}
    \centering
    \includegraphics{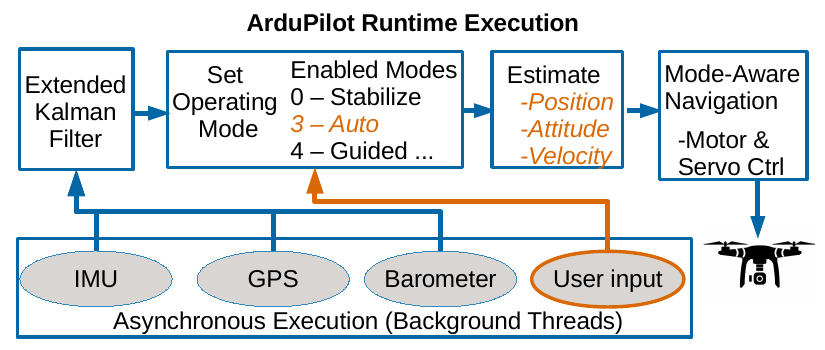}
    \caption{ArduPilot uses support for multiple threads to query sensors and asynchronously update its main control loop. }
    \label{fig:draw-ArduPilot}
\end{figure}

The rest of the paper is organized as follows.
\S\ref{sec:sensorbugs} provides background on
sensor bugs and UAVs. \S\ref{sec:bug-study} shows the impact of sensor bugs on UAV reliability.
\S\ref{sec:Avis} presents the design of Avis,
including our framework to create workloads that
exercise operating modes and our stratified
approach for fault injection.
\S\ref{sec:implementation} describes
implementation details.
In \S\ref{sec:evaluation}, we
present our evaluation of Avis and a study 
analyzing sensor bugs and their manifestations in
UAVs. We discuss related work in
\S\ref{sec:related-work}. Finally, we conclude in
\S\ref{sec:conclusion}.

\section{Background}
\label{sec:sensorbugs}
Figure~\ref{fig:draw-ArduPilot} depicts runtime 
execution for ArduPilot, a widely used software
system for controlling UAVs~\cite{ArduPilot}.  Created in 2007 by
hobbyists, ArduPilot is now used by more than 65
companies in industrial applications.  It supports a wide
range of aircraft from large fixed-wing planes 
to copters that weigh less than a kilogram.
The code base now exceeds 700K lines with nearly
100 developers contributing to its maintenance.
PX4 is another popular, open-source framework for
autopilot control firmware~\cite{px4}.  PX4 has
over 6M lines of code and is used by the
production-grade PixHawk UAV.

As shown in Figure~\ref{fig:draw-ArduPilot}, 
ArduPilot uses multiple parallel threads to 
read from sensors and manage flight dynamics.
Pilots can provide input with a remote control 
or with a laptop.  
Throughout this paper, we refer to pilot inputs
as the UAV workload, i.e., a sequence of flight 
commands.  For example, ArduPilot supports 
flight commands to (1) directly adjust thrust, 
yaw or pitch, and (2) fly to a waypoint coordinate.
The code used to execute these commands differs.
An {\em operating mode} encompasses all code 
execution associated with a pilot command.
Today, the ArduPilot code base supports 25 
operating modes including takeoff,
landing, manual piloting, fly to waypoint, 
return home, auto avoidance and acrobatics.
In addition, developers can add custom 
modes to create automated flight maneuvers.

During every iteration of the simulation, an operating mode 
translates user inputs and sensor signals
to actuation in the motor systems.  To help 
developers, ArduPilot includes models to 
estimate the state of the aircraft.  For any 
operating mode, it is important to know the 
position, altitude and attitude of the aircraft
before adjusting motor systems.  However, 
sensor failures can render these models useless 
because (1) fault handling logic may not 
realize that state models differ from 
normal flight conditions (as in Figure~\ref{fig:draw-bugexample})  
and (2) sensor failures may lead to incorrect
state models that diverge from reality.
When developer expectations, state models 
and reality differ, the UAV is flying in 
an unsafe condition that could have serious consequences.

\begin{figure}
    \centering
    \includegraphics{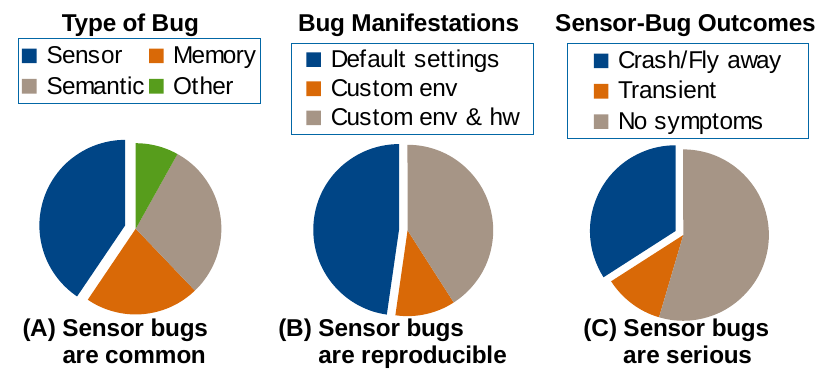}
    \caption{Analysis of reported bugs for ArduPilot and PX4.}
    \label{fig:draw-bugstudy}
\end{figure}

\section{The Impact of Sensor Bugs}\label{sec:bug-study}
We reviewed bugs reported and resolved on the 
public GitHub repositories of ArduPilot (206 cases) 
and PX4 (188 cases) from 2016-2019.  
In total, we reviewed 394 bugs.  We excluded 
bug reports related to software development 
environments and tools (29).  We also removed duplicates, false reports,
reports unrelated to control firmware and bugs that
were described too vaguely to repeat or
understand (150).  After pruning, we were left with 215
bugs.

We classified bugs by their root causes:
\emph{Semantic} bugs were caused by 
logically incorrect behavior of the UAV without 
a preceding hardware fault; \emph{Memory} bugs
stemmed from incorrect memory allocation or invalid
accesses; \emph{Sensor} bugs, as described earlier, 
were triggered by a sensor fault.  Finally, we 
grouped all remaining bugs, including concurrency 
bugs, under the label \emph{other}.

We also classified bugs by the flight
conditions where they manifested.
Some bugs were easy to reproduce, because
they could be triggered under default 
settings, i.e., with standard environment
and hardware configurations.  We 
distinguished bugs that required special settings.
Finally, we also classified bugs by their 
symptoms.  Some bugs were asymptomatic. 
Others had transient affects, such as jerks
during flight.  The most serious bugs 
resulted in a crash or the UAV flew away.

\begin{center}
 \vspace{-7pt}
  \begin{tabular}{|p{.95\columnwidth}|}
  \hline
  \textit{\textbf{Finding 1: Sensor bugs account for 20\% of control firmware bugs.}}\\
  \hline
  \end{tabular}
\end{center}

We found that semantic bugs accounted for 
68\% of reported bugs.  Sensor bugs were 
second most common, accounting for 20\% of
reported bugs.  However, as shown in Figure~\ref{fig:draw-bugstudy}(A),
sensor bugs represented 40\% of reported 
bugs that caused the UAV to crash. 

We believe sensor bugs are common for a several reasons.
(1) ArduPilot and PX4 have adopted 
Valgrind \cite{valgrind} to detect memory bugs during in-house testing \cite{ArduPilot-valgrind, px4-valgrind},
depressing the frequency of such bugs.
(2) Semantic bugs were often asymptomatic (90\%).
Common symptoms include improper messages 
appearing in logs, unimplemented commands, 
gradual vehicle drift, and mishandled unit 
conversions resulting in the vehicle navigating to an incorrect location.
These bugs reflected the growing number of 
contributors to the code base, but they were  
resolved without serious consequences.
(3) As of this writing, ArduPilot and PX4 did
not adopt tools for rigorous fault injection.
Code paths related to handling sensor failures
are checked by unit tests, but are not comprehensively 
checked across multiple environments.

\begin{center}
  \vspace{-7pt}
    \begin{tabular}{|p{.95\columnwidth}|}
    \hline
    \textit{\textbf{Finding 2: 47\% of the sensor bugs are reproducible under default settings.}}\\
    \hline
    \end{tabular}
\end{center}

Figure~\ref{fig:draw-bugstudy}(B) examines the
44 sensor bugs in our study.  47\% could be 
reproduced under default environment and hardware
settings.  In a nutshell, these bugs followed a 
simple template (1) trigger a sensor failure and 
(2) check the vehicle's behavior for symptoms.
Wind and humidity contributed to bugs that 
required special environmental conditions.
However, new aircraft also introduced bugs.
For example, \px{12758} in PX4 describes
a sensor bug where the fault handling logic
in a new copter used the wrong interface to 
set return to home mode on the aircraft.

\begin{center}
  \vspace{-7pt}
  \begin{tabular}{|p{.95\columnwidth}|}
  \hline
  \textit{\textbf{Finding 3: About 34\% of the sensor bugs have serious symptoms.}}\\
  \hline
  \end{tabular}
\end{center}

Not only were sensor bugs the most common
root cause for bugs manifesting as a crash,
Figure~\ref{fig:draw-bugstudy}(C) shows that 
most reported sensor bugs displayed symptoms.
A significant portion of sensor bugs are serious (34\%).
This finding demonstrates the importance of detecting sensor bugs in RVs.
Sensor bugs are prone to serious outcomes 
because UAV depend on sensors for safe flight.
We noticed that, for many root 
causes, developers applied default actions, 
like return to home, assuming they can be
executed effectively.  When sensors failed,
these assumptions--i.e., the difference
between expectations, modeled state and 
reality--- had severe consequences.

\section{Avis Design}
\label{sec:Avis}

\begin{figure}
    \centering
    \includegraphics[width=\columnwidth]{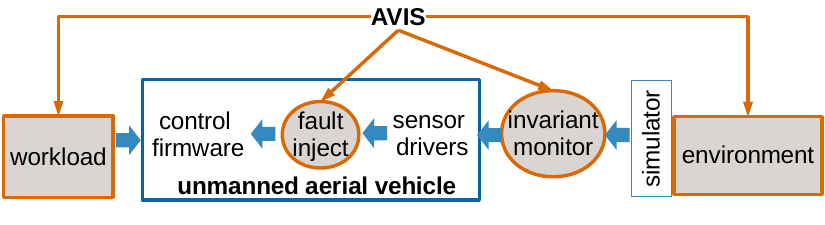}
    \caption{An overview of Avis. Arrows denote the direction of information flow.}
    \label{fig:Avis_overview}
\end{figure}

As shown in Figure \ref{fig:Avis_overview}, Avis consists of three major components: workloads, a fault injection engine, and an invariant monitor. Avis tests a UAV by simulating its behavior in a physical \emph{environment} under a workload. Workloads issue flight commands to the UAV, as shown in Figure \ref{fig:Avis_overview}. While the UAV runs, the fault injection engine monitors the vehicle's mode transitions. The fault injection engine uses mode transitions to schedule injections. Meanwhile, the invariant monitor checks the UAV's simulated physical state to detect unsafe conditions. If an unsafe condition occurs, the invariant monitor generates a detailed report to help reproduce and diagnose the bug.

UAV simulation involves executing mostly unmodified UAV source code while simulating hardware. The only two modifications are the use of simulated sensor and actuator drivers.  The sensor drivers read from the simulator instead of hardware. The original firmware source code uses the simulated sensor inputs to determine its next motor controls. The actuator drivers communicate the motor controls to the simulator (not shown). The simulator uses these controls to generate the vehicle's new physical state. One iteration of this communication is called a simulation \emph{time-step}.

Avis relies on simulation instead of real UAV flights for three reasons. First, recall that sensor bugs can have serious symptoms; simulating the behavior of the UAV under a fault injection scenario allows Avis to expose a sensor bug without suffering from the bug's symptom, e.g., crashing the vehicle. Second, simulations can be performed faster than real experiments, improving test throughput. Last but not least, all UAV firmware modules (except for drivers) are identical to the ones used in real systems, enabling Avis to use simulation to check real UAV firmware.
Next, we discuss the three components of Avis in more details.

\subsection{Workloads and Environments}
\label{subsec:design-workloads}
Pilots send commands using a ground-control station to control a UAV's movements. A sequence of pilot commands constitutes a workload. UAVs typically communicate with the ground-control station using the MAVLink \cite{mavlink} protocol. Ideally, all control firmware would support the same MAVLink messages and strictly implement their semantics. In practice, implementations have subtle quirks that make it difficult for users to develop portable workloads. To mitigate this issue, Avis provides default workloads that work on both ArduPilot and PX4. We also provide a high-level framework developers can use to extend our workloads and build their own.

We design our workloads to exercise common commands such as takeoff, fly-to-waypoint, and land. Each command maneuvers the vehicle in a simple way, e.g. along a polygon. This allows Avis to trigger bugs that UAV pilots are most likely to experience.

The simulator provides an \emph{environment}, a model of the physical world that contains obstacles and weather effects. Workloads navigate the UAV in the environment. Some unsafe conditions can only be recreated in specific environments, e.g. due to adverse weather or obstacles such as trees. Avis uses an environment without hostile weather or obstacles. 

\subsection{Fault Injection Engine}
\label{subsec:injecting-sensor-faults}

Avis injects sensor failures during simulated flights to expose bugs
in control firmware that lead to unsafe conditions.
The main challenge Avis faces is exploring a huge fault space.  
Exhaustively injecting every possible fault is not feasible and,
in most cases, would yield normal executions that do not aid
root cause analysis.  In this subsection, we first elaborate on 
this challenge.  Then, we propose a new search strategy, called SABRE,
for fault injection based on the UAV's operating mode. 
Finally, we show how to avoid fault injections that yield redundant 
states to further improve search efficiency.  

\vspace{0.1in}
\noindent \textbf{Fault Model and Challenges:} Avis models \emph{clean sensor failures}, where a sensor instance stops communicating with the firmware and the driver reports the instance has failed. Any sensor instance can fail at any time (controlled by Avis). Moreover, a failed sensor will not recover during the same test run. Avis focuses on such a simple fault model because it is realistic. More importantly, UAVs are expected to handle this simple fault model.




\begin{figure}
    \centering
    \includegraphics[width=\columnwidth]{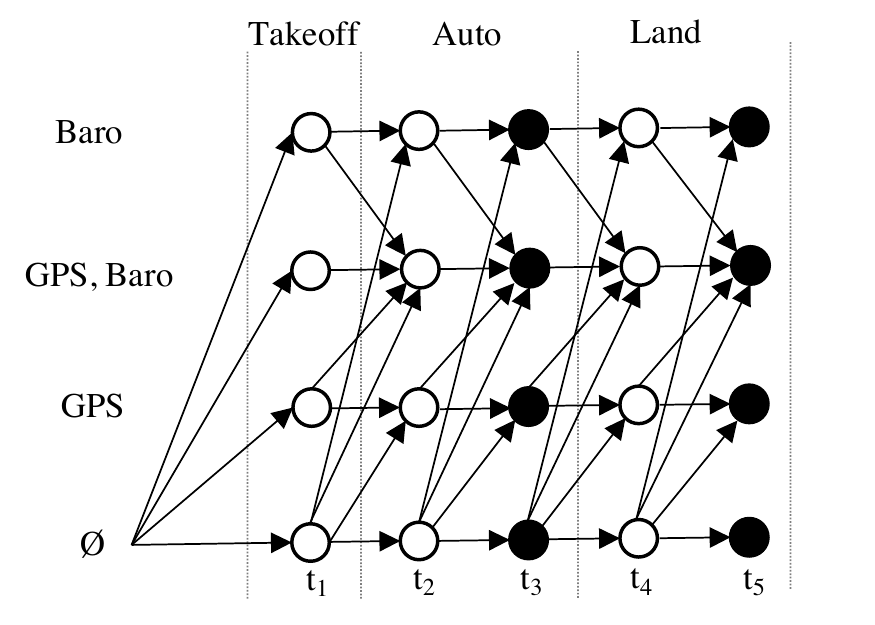}
    \caption{UAV modes and the corresponding UAV code executed at different times $t_i$ during a test run. Each circle represents the failure state of the GPS and barometer. Similar states are colored black.}
    \label{fig:st_xplod}
\end{figure}

Usually, a UAV samples its sensors thousands of times each second. Consequently, there are far too many fault injection sites to exhaustively cover. Moreover, since UAV workloads usually take minutes to execute, effectively exploring fault injection sites becomes even more important. On a simple vehicle with 7 onboard sensors and no backups, there are more than $(2^7 - 1) \times 10^3 \approx 10^5$ fault injection sites \emph{each second}.

To maximize the number of unsafe scenarios identified as we search the fault space, we rely on a key observation: \textbf{there are many similar fault injection sites within each mode}. Figure \ref{fig:st_xplod} demonstrates this observation using an example. Since injecting the same failures in the same mode likely leads to the same UAV behavior, injecting sensor failure at $t_3$ can be similar to injecting failures at $t_2$. However, injecting sensor failures at $t_4$ exposes different UAV code to failures, likely causing different UAV behaviors.

This observation motivates us to {prioritize fault injection at mode boundaries.} Consider the bug described in Figure \ref{fig:draw-bugexample}. In a narrow window while the UAV has low altitude but has not yet landed, it is vulnerable to an IMU failure blinding the firmware to the effects of its own actuation. By considering the area between landing and disarming early in the injection space exploration, Avis quickly triggers this scenario.

\vspace{0.1in}
\noindent {\bf Search Strategies:}
Inspired by the previous observation, we propose \emph{SABRE---a \underline{s}tr\underline{a}tified \underline{bre}adth-first search}.
SABRE explores the space of sensor failures using injection sites across all modes. Before we describe SABRE, 
we first consider two common search strategies to understand their drawbacks.

Figure \ref{fig:st_xplod} shows the fault space that should be explored given two sensors (GPS and Barometer) and a workload with five time-steps. Depth-first search is an intuitive way to search the fault space, which results in the following sequence of executions: 

\begin{center}
$\left<\emptyset, \emptyset, \emptyset, \emptyset, \emptyset\right>$\\
$\left<\emptyset, \emptyset, \emptyset,  \emptyset, \{\mathbf{GPS}\}\right>$\\
$\left<\emptyset, \emptyset, \emptyset, \emptyset, \{\mathbf{Baro}\}\right>$\\
$\left<\emptyset, \emptyset, \emptyset, \emptyset, \{\mathbf{GPS}, \mathbf{Baro}\}\right>$\\
$\left<\emptyset, \emptyset, \emptyset, \{\mathbf{GPS}\}, \{\text{GPS}\}\right>$\\
$\dots$
\end{center}

\noindent
In each sequence $\left<F_1, \ldots, F_5\right>$, $F_i$ denotes the set of sensors that are failed at time $t_i$. This search is ineffective because similar fault injection scenarios (e.g. failing the GPS at $t_4$ and failing the GPS at $t_5$) are explored before the scenarios in different modes (e.g., failing GPS at $t_3$). Given a limited test budget, depth-first search tends to test a small area of the UAV firmware. 

An alternative approach is to use breadth-first search to explore the fault space. We explore the fault space across time to reach dissimilar moments faster. This approach results in the following sequence of executions:

\begin{center}
$\left<\emptyset, \emptyset, \emptyset, \emptyset, \emptyset\right>$ \\
$\left<\{\mathbf{GPS}\}, \{\text{GPS}\}, \{\text{GPS}\}, \{\text{GPS}\}, \{\text{GPS}\}\right>$ \\ 
$\left<\{\mathbf{Baro}\}, \{\text{Baro}\}, \{\text{Baro}\}, \{\text{Baro}\}, \{\text{Baro}\}\right>$ \\
$\left<\{\mathbf{GPS}, \mathbf{Baro}\}, \{\text{GPS}, \text{Baro}\}, \ldots, \{\text{GPS, Baro}\} \right>$ \\
$\left<\emptyset,  \{\mathbf{GPS}\}, \{\text{GPS}\}, \{\text{GPS}\}, \{\text{GPS}\}\right>$ \\
$\dots$ \\
$\left<\emptyset, \emptyset, \{\mathbf{GPS}\}, \{\text{GPS}\}, \{\text{GPS}\}\right>$ \\
\ldots
\end{center}

\noindent
However, this strategy is ineffective because it also explores similar fault scenarios first. Specifically, after injecting failures at $t_2$, breadth-first search considers similar failures at $t_3$ next. This delays exploration of complex fault scenarios (i.e., failing different sensors at different times) until all the simple scenarios are checked. Given the limited test budget, complex fault scenarios may never be explored by breadth-first search.

In contrast to depth-first search and breadth-first search, SABRE prioritizes exploring the most different states in the fault space by considering the UAV's mode. Specifically, SABRE first explores the scenarios that inject sensor failures around mode transitions, allowing SABRE to consider fault scenarios that fail different sensors at different modes before the aforementioned two strategies. Note that SABRE only prioritizes the search to uncover bugs earlier -- exhaustive search is still possible, but is prohibitively expensive.

\begin{algorithm}[t]
\LinesNumbered
\SetAlgoLined
\SetKwInOut{Workload}{Workload}
\SetKwInOut{Failures}{Failures}
\Workload{the workload to execute}
\Failures{the sensor failures to inject}
transitionQueue $\leftarrow$ Queue(ProfileExperiment(Workload))\;
seenBugs $\leftarrow$ \{\}\;
\While{transitionQueue is not empty} {
    timestamp, injectedFailures $\leftarrow$ Dequeue(transitionQueue)\;
    \For{failureSet in PowerSet(Failures)}{
        \If{CanPrune(timestamp, failureSet, seenBugs, injectedFailures)} {
            continue\;
        }
        failures $\leftarrow$ injectedFailures $\cup$ \{(failure, timestamp) : failure $\in$ failureSet\}\;
        result $\leftarrow$ RunExperiment(Workload, failures)\;
        \eIf{Ok(result)} {
            \For{modeTimestamp $\in$ result.modeTransitions}{
                Enqueue(transitionQueue, (modeTimestamp, failures))\;
            }
        } {
            reportBug(failures, result)\;
            seenBugs $\leftarrow$ seenBugs $\cup$ \{failures\}\;
    }
}
    Enqueue(transitionQueue, (timestamp + 1, injectedFailures))\;
}
\caption{SABRE}
\label{alg:Avis}
\end{algorithm}

Algorithm \ref{alg:Avis} shows how Avis uses SABRE to guide its fault-space exploration. Here, we walk through the algorithm using the example shown in Figure \ref{fig:st_xplod}. Avis first executes the workload to determine when mode transitions occur (Line 1). Mode transitions are discovered at $t_1,$ $t_2,$ and $t_4$. As a result, Avis initializes its transition queue to $\left<(t_1, \emptyset), (t_2, \emptyset), (t_4, \emptyset)\right>$, where each $(t_i, set)$ means to inject new faults at $t_i$ alongside the fault combinations $\left<sensor, timestamp\right>$ in $set$. Next, Avis dequeues the injection scenario $(t_1, \emptyset)$ from the queue (Line~4) and applies all possible sensor failures to this point (Line 5) but only if they are not redundant (Lines 6-8). Thus, Avis tests the following executions:

\begin{center}
$\left<\{\mathbf{GPS}\}, \{\text{GPS}\}, \{\text{GPS}\}, \{\text{GPS}\},  \{\text{GPS}\}\right>$\\
$\left<\{\mathbf{Baro}\}, \{\text{Baro}\}, \{\text{Baro}\}, \{\text{Baro}\},  \{\text{Varo}\}\right>$\\
$\left<\{\mathbf{GPS}, \mathbf{Baro}\}, \{\text{GPS, Baro}\}, \ldots, \{\text{GPS, Baro}\}\right>$\\
\end{center}
 
\noindent
Avis also re-enqueues each bug-free scenario it tests
for generating new fault scenarios in later runs (Lines 11-14). Finally, Avis re-enqueues the dequeued scenario with a changed timestamp so that it will explore injecting faults at different times in later runs.
The next tuple dequeued by Avis is $(t_2, \emptyset)$ since it is the second mode transition discovered during the profiling run. So, Avis injects faults at $t_2$ as it did at $t_1$. Next, Avis dequeues the mode transition $(t_4, \emptyset).$ So, rather than conducting fault injection at $t_3$ next like breadth-first search, Avis considers this fault combination:
\begin{center}
$\left<\emptyset, \emptyset, \emptyset, \{\mathbf{GPS}\}, \{\text{GPS}\}\right>$.\\
\end{center}

\noindent
In this way, Avis prioritizes injecting faults around the mode transitions.
This process repeats until the queue is exhausted.


\subsubsection{Redundancy Elimination} 
\label{subsec:redunduncies}

\begin{figure*}[h!t]
\centering
    \begin{subfigure}[t]{0.19\textwidth}
        \centering
        \includegraphics[scale=.6]{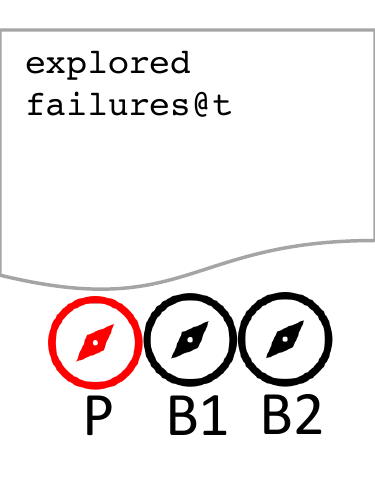}
        \caption{}
        \label{fig:syma}
    \end{subfigure}
    \begin{subfigure}[t]{0.19\textwidth}
        \centering
        \includegraphics[scale=.6]{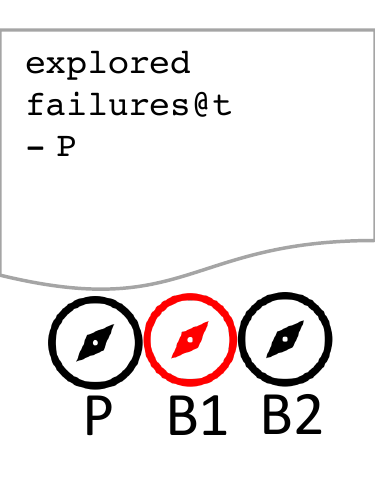}
        \caption{}
        \label{fig:symb}
    \end{subfigure}
    \begin{subfigure}[t]{0.19\textwidth}
        \centering
        \includegraphics[scale=.6]{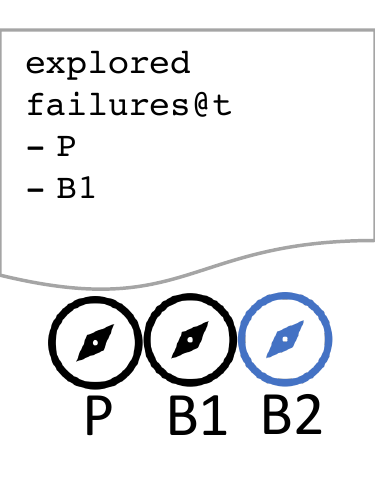}
        \caption{}
        \label{fig:symc}
    \end{subfigure}
    \begin{subfigure}[t]{0.19\textwidth}
        \centering
        \includegraphics[scale=.6]{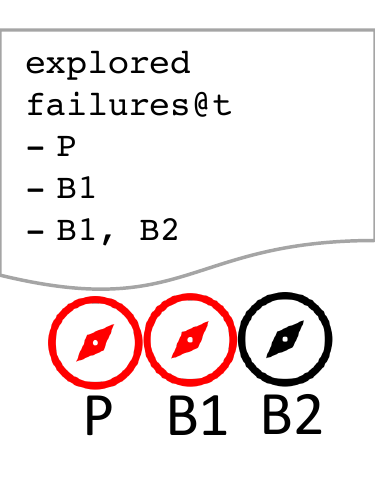}
        \caption{}
        \label{fig:syme}
    \end{subfigure}
    \begin{subfigure}[t]{0.19\textwidth}
    \centering
    \includegraphics[scale=.6]{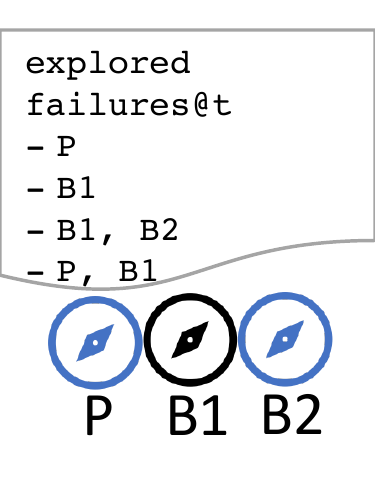}
    \caption{}
    \label{fig:symf}
    \end{subfigure}
        \caption{The process of pruning redundant faults. Compasses colored red are candidates for fault injection; colored black are not under current consideration; colored blue are considered but pruned. Not shown: $\left\{\text{P, B1, B2}\right\}$ and $\left\{\text{B1, B2}\right\}$.}
    \label{fig:sym}
\end{figure*}
While SABRE guides the order that injection sites are searched, it does not avoid redundant injection scenarios. Avis uses two policies, i.e., \emph{found bug pruning} and \emph{sensor instance symmetry}, to eliminate these redundancies.


In the {found bug pruning} policy, if injecting a sensor failure $F_0$ at time $t$ triggers a bug, Avis will not try to inject $F_0$ \emph{plus} other failures at time $t$ in the later test runs. The intuition behind this policy is that if a vehicle cannot handle a single sensor failure then it is unlikely to correctly handle multiple failures in the same program context.

The {sensor instance symmetry} policy exploits the role of a sensor, i.e., primary or backup, to reduce the combinatorial size of the fault space. UAV systems are usually equipped with multiple sensor instances of the same sensor type to tolerate sensor failures. One of these redundant sensors is the primary, while the other instances are the backups. We find that, when handling sensor failures, the UAV's behavior depends on the role of the failed sensors instead of which instances fail. Therefore, Avis skips a sensor failure scenario if the same failed sensor roles have been tested before, regardless of the actual instances.

Figure \ref{fig:sym} illustrates the sensor instance symmetry policy with an example. Consider a UAV with three compasses labeled as ``P,'' ``B1,'' and ``B2,'' corresponding to the primary and the two backups respectively. Assume Avis is injecting sensor failures at time $t$. In the first two runs, Avis fails sensor P (Figure \ref{fig:syma}) and sensor B1 (Figure \ref{fig:symb}), respectively. These are two different scenarios since P is a primary sensor while B1 is a backup sensor. Then, Avis considers failing B2 but decides to skip it (Figure \ref{fig:symc}). This is because B2 is a backup sensor and Avis has tried failing one backup sensor in a previous run (Figure \ref{fig:symb}). Later, Avis injects failures at ``P'' and ``B1'' simultaneously, since it has not yet injected a failure of a primary with a backup (Figure \ref{fig:syme}). When Avis considers injecting a failure of ``P'' and ``B2'' in Figure \ref{fig:symf}, it sees that it has already failed ``P'' and ``B1'' and skips this combination.


In general, if a vehicle is equipped with $N$ instances of a sensor, sensor instance symmetry reduces the number of faults that must be injected from $N \times (2^N - 1)$  (e.g. $N$ primary instances of a sensor with $2^N$ subsets to fail minus the empty set of failures) to $2N - 1$ (e.g. $N - 1$ ways to fail the backup sensors with or without the primary, plus one way to fail the primary alone) thus diminishing the effects of state explosion. For the above example, sensor instance symmetry reduces the number of checks from 21 to 5.

\subsection{Invariant Monitor}
\label{subsec:invariants}
At the end of each simulation iteration, Avis's invariant monitor checks two simple rules:
\begin{itemize}
    \item \textbf{Safety} - The UAV does not collide with an obstacle.
    \item \textbf{Liveliness} - The UAV must always make progress towards its goal. This may be compromised under special circumstances to preserve safety.
\end{itemize}

\subsubsection{Safety}
The invariant monitor detects both software crashes and physical collisions for the safety rule. To detect software crashes, the invariant monitor checks if the firmware process is still running. To detect physical collisions inside the simulator, the invariant monitor checks if (1) the vehicle rapidly (de)accelerates but (2) has the same position as another simulated object, e.g, the ground.

\subsubsection{Liveliness}
Checking the liveliness condition is challenging for two reasons. First, the behavior of the UAV may change in the presence of sensor failures or non-determinism introduced by the operating system scheduler (e.g. slight delays between the workload sending and the firmware receiving messages), although the mission is still correctly being executed. Second, liveliness sometimes must be sacrificed in the presence of sensor failures to preserve safety. Avis must detect when this has occurred and not report an error.

To combat the first issue, Avis detects liveliness violation by measuring the differences in the UAV's behavior between the test run and a set of correct profiling runs. If the test run significantly diverges from the correct runs, then liveliness is violated. We assume runs without sensor failures are correct. 

To measure the difference between two runs, Avis compares the states of the vehicle at the same time offset $t$ in both runs. The state of the vehicle is represented using the tuple $(P, \alpha, M)$, where $P \in \mathbb{R}^3$ is the vehicle's position, $\alpha \in \mathbb{R}^3$ is the vehicle's acceleration, and $M$ is the vehicle's mode. Velocity is excluded because it is redundant: if the difference in velocity is large, then the difference in acceleration or position must also be large. We could detect liveliness violations using position alone. However, it takes tens of seconds to detect liveliness violations with this approach. Using multiple variables lets us detect violations in seconds. The invariant monitor reports a liveliness violation if the state in the test run deviates from the states in the profiling runs.

Before defining the distance between two states, we first define the distance of each component in the state tuple. For both the position $P$ and the acceleration $\alpha$, we use Euclidean distance ($d_e$). For example, the distance between two positions $P_1$ and $P_2$ is computed as
$$d_e(P_1, P_2) = \sqrt{(P_1^x - P_2^x)^2 + (P_1^y - P_2^y)^2 + (P_1^z - P_2^z)^2},$$
where $P_i^x$, $P_i^y$, and $P_i^z$ are the three coordinate values of $P_i$.
To define the distance between two modes, we utilize the \emph{mode graph}.
A mode graph is a directed graph, where each node represents a mode and each edge represents a \emph{mode-change} event. The mode graph is constructed from the observed transitions between modes in the profiling runs. Note that not every mode is adjacent in the mode graph -- for instance, a drone cannot land before it is flying. The distance between modes (denoted $d_m$) is defined as the length of the shortest path between them in the firmware's mode graph. 

We also normalize the distance of each component before computing the distance between two states. Intuitively, we want to transform the distance between the acceleration and position components to measure ``on a scale from 0 to the longest path in the mode transition graph, how far apart are these values?'' 
To normalize the distance on positions, we first compute $\mathcal{P}$, the largest distance between any two positions that occur at the same time $t$ of two different runs. Let $P_{i, t}$ denotes the position of the vehicle in simulation $i$ at time $t$. Then, $\mathcal{P}$ can be computed as
$$\mathcal{P} = \text{max}\left\{d_{e}(P_{i, t}, P_{j, t}) | 1 \leq i, j \leq N \land 1 \leq t \leq T\right\},$$
where $N$ is the number of profiling runs and $T$ is the duration of the profiling runs. 
To ensure that every profiling runs have the same duration, we repeat the last state an appropriate number of times for the shorter runs.
Then the normalized position distance can be computed as
\[d_P(P_{i, t}, P_{j, t}) = \frac{d_e(P_{i, t}, P_{j, t}) \mathcal{D}}{\mathcal{P}},\]
where $\mathcal{D}$ denotes the length of the longest path in the mode graph. Similarly, the normalized distance of two accelerations can be computed as
\[d_A(A_{i, t}, A_{j, t}) = \frac{d_e(A_{i, t}, A_{j, t}) \mathcal{D}}{\mathcal{A}},\]
where $A_{i, t}$ denotes the acceleration of the vehicle in simulation $i$ at time $t$, and
$$\mathcal{A} = \text{max}\left\{ d_{e}(A_{i, t}, A_{j, t}) | 1 \leq i, j \leq N \land 1 \leq t \leq T\right\}$$ 
is the largest distance between any two accelerations at the same time $t$ of two different runs.

Finally, the distance between two state tuples is defined as
\[d(S_{i, t}, S_{j, t}) = \norm{(d_P(P_{i, t}, P_{j, t}), d_A(A_{i, t}, A_{j, t}), d_M(M_{i, t}, M_{j, t}))}\]
where $M_{i, t}$ denotes the mode of the vehicle at time $t$ in simulation $i$ and $\norm{.}$ denotes the Euclidean norm.


With this distance defined, we can compute $\tau$, the largest distance between any two states at the same time $t$ of two different runs to be
\[\tau = max\left\{d(S_{i, t}, S_{j, t} | 1 \leq i, j \leq N \land 1 \leq t \leq T)\right\}.\]
Avis considers the liveliness to have been violated in simulation $S_F$ if
\begin{equation}
    \label{invariant:liveliness}
    \forall i: d(S_{F, t}, S_{i, t}) > \tau. 
\end{equation}
That is, liveliness is violated at time $t$ if the state is further from all profiling runs than the maximum seen  distance.

To allow UAVs to preserve safety at the expense of liveliness, we allow developers to specify \emph{safe modes} that are always permitted. For instance, we provide a safe return to launch location mode. If a vehicle enters a safe mode, Avis does not signal that a bug has been found, even if liveliness has been violated. Additional invariants must be supplied for safe modes. For example, a vehicle executing in the return to launch mode must make progress back to the launch site. 

\subsection{Replaying Bugs}

Avis records the failures that it injects. Avis saves the failures for replay if an unsafe condition is found. To reconstruct the unsafe condition, Avis re-executes the mission, injecting the same faults at the same time \emph{offsets} from mode transitions. Even in the presence of minor non-determinism this technique is successful since failures are injected at the same time relative to the modes they affect.

\section{Implementation}
\label{sec:implementation}
Avis contains several components: (1) a high-level framework for building UAV workloads, (2) a fault injection engine for generating fault injection scenarios, and (3) an invariant monitor for detecting incorrect firmware behaviors. Avis' source code is available at \cite{avissource}. 

\begin{figure}[t]
    \centering
    \includegraphics[width=\columnwidth]{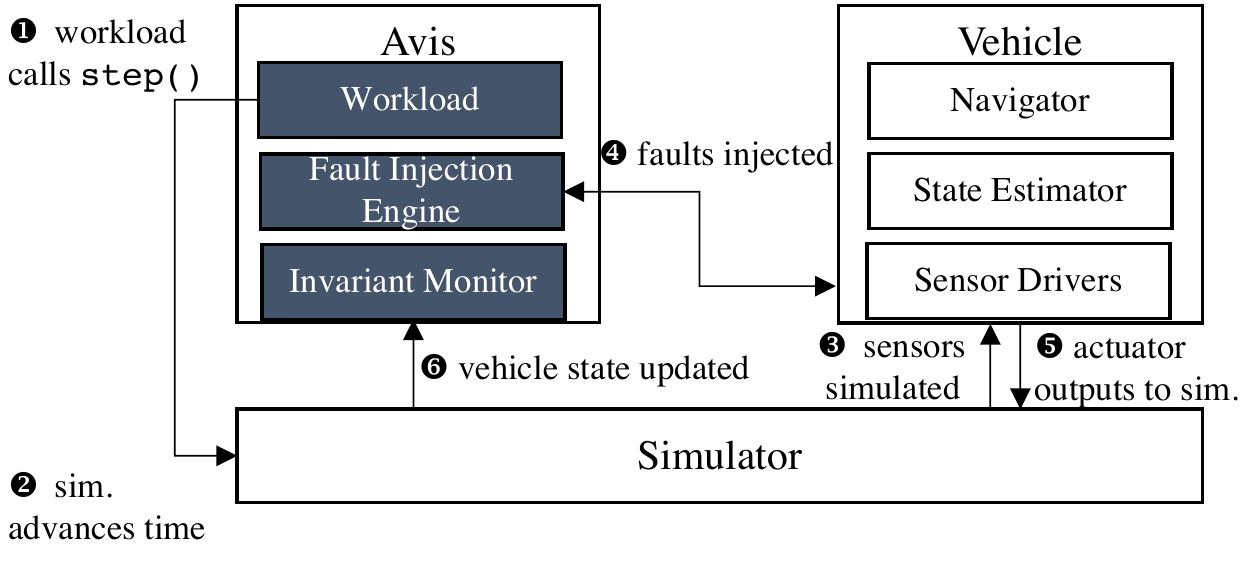}
    \caption{An overview of one step in the Avis process.}
    \label{fig:Avis-process}
\end{figure}

Figure \ref{fig:Avis-process} shows an overview of one time-step of the simulator. The goal of Avis is to test the UAV's firmware under different sensor failure scenarios. At the start of each test, Avis provisions a new instance of the simulator and firmware. Next, Avis launches its invariant monitor and its workload. The workload executes until it returns control back to Avis by calling the \code{step()} RPC (Step~1). Typically, workloads created using our framework only need to call our high-level APIs, e.g., \code{takeoff()}, which call \code{step()}. When \code{step()} is called, Avis notifies the simulator to advance its time by a fixed unit (1ms) and to permit the firmware to retrieve its current state (Step~2). The simulator then generates synthetic sensor readings based on the UAV's physical state (Step~3). After reading the sensor values, instrumented code in sensor drivers report the readings to the fault injection engine and inject sensor failures as directed (Step~4). Then the firmware continues executing and eventually sends the actuator outputs to the simulator (Step~5). The simulator uses this information to compute the next physical state of the vehicle (Step~6) and it informs Avis the step has completed. At the end of each step, the invariant monitor checks the vehicle's state is safe.

\subsection{Workload Framework and Workloads}


UAVs communicate a workload's commands using the MAVLink protocol \cite{mavlink, pymavlink}. 
However, MAVLink is challenging for developers to use to create workloads. The UAV is responsible for controlling most interactions between the ground-control station and the vehicle. For instance, to upload new missions the ground-control station first communicates the number of mission items to the vehicle and then waits for the vehicle to request each item. This presents two problems. First, it introduces the possibility of deadlock during model checking. Since the vehicle's execution is carefully synchronized with both the simulator and Avis, both parties must avoid simultaneously waiting on messages from each other. Second, this makes even simple missions difficult to implement. 

Avis's workload framework provides high-level APIs that safely abstract the most common MAVLink transactions. By default, we provide two workloads that we show are effective at triggering bugs with Avis. Developers can create additional workloads using our Python framework. 

We show an example of a simple workload that uses our framework in Figure \ref{fig:example_workload}. First, the workload waits 40 seconds for the UAV to initialize (Line 1). Next, the workload uploads takeoff and land commands (Line 2). Then, the workload arms the UAV (Line 8) and enters the fly-to-waypoint mode (Line 9). Finally, the workload waits for the vehicle to reach its target altitude (Line 10) and then for the vehicle to land (Line 11). The final step of the workload is to communicate the test succeeded to AVIS (Line 12).

\begin{figure}
\begin{lstlisting}
class AutoWorkload(workload_framework.Target):
    def test(self):
        self.wait_time(40000) 
        self.upload_mission(
            self.takeoff_mission(20,self.cur_lati,
                self.cur_longi,self.home_alti) + 
            self.land_mission())
        self.arm_system_completely()
        self.enter_auto_mode()
        self.wait_altitude(20)
        self.wait_altitude(0)
        self.pass_test()
\end{lstlisting}
\caption{An example workload built with Avis's workload framework.}
\label{fig:example_workload}
\end{figure}

Our first workload uses a manual mode that holds the vehicle's position. First, the UAV ascends to an altitude of 20 meters (m). Then, the UAV flies the perimeter of a 20mx20m box. Finally, the UAV lands at its launch point. 

This mission is sufficient to test manual modes. Other manual modes maintain the vehicle's orientation (e.g. pitch/roll/yaw) or altitude. Holding the position requires holding orientation and altitude. UAV firmware typically reuses the code that implements this behavior. So, by testing the position mode, we test these two modes as well. 

UAV firmware also provide stunt and race modes. We choose to leave these modes untested. Stunts and race modes relax the firmware's safety guarantees. This places more trust in the UAV's operator. A sensor failure at this time cannot expose a new bug.

Our second workload uses waypoints and a fence. Fences are used to prevent the UAV from entering restricted airspace. Fences can also be used to contain a UAV. First, the UAV ascends to an altitude of 20m. Then, our workload guides the UAV along a 20mx20m box. The box overlaps with a fenced area the UAV must avoid. The UAV lands at its launch site.

We do not consider the effect of special workload details or environments on bug manifestation. We observe that known sensor bugs are not sensitive to these factors. Nevertheless, future work may rigorously pursue this direction to establish the absence of this class of bugs.

\subsection{Fault Injection Engine}

Avis's fault injection system is composed of two components. The first component is \code{libhinj} (\textbf{H}ardware Fault \textbf{Inj}ector), a library for instrumenting UAV firmware. The second component is the scheduler. The scheduler injects failures by communicating with simulated drivers instrumented with \code{libhinj}. Here, we discuss (1) the implementation of \code{libhinj} and (2) the implementation of the scheduler.

\subsubsection{libhinj}
We implement \code{libhinj}, a library that functions as the interface between Avis and the UAV firmware. \code{libhinj} reports the firmware's mode transitions and sensor readings to Avis and injects sensor failures. \code{libhinj} is available at \cite{libhinj}.

\code{libhinj} reports the firmware's mode to Avis through its \code{hinj\_update\_mode()} API. UAV firmware has a specific function that updates the vehicle's mode. We simply insert the \code{hinj\_update\_mode()} call within this call site. As a result, whenever the mode changes, \code{hinj\_update\_mode()} is invoked to report the updated mode to Avis.

We use \code{libhinj} to instrument the firmware's driver module. This allows Avis to inject faults on-demand. We insert a \code{libhinj} API call in the \code{read()} procedure of each sensor driver. The API call queries the scheduler to determine if the read should fail. The API call returns the scheduler's decision. If the sensor should be failed, the API overwrites the sensor reading and the instrumented code executes the firmware's error-handling code. \code{libhinj} supports fault injection for various types of sensors including gyroscopes, accelerometers, GPSs, compasses, and barometers.

To facilitate testing UAVs, we integrate \code{libhinj} into two dominant open-source UAV firmware, ArduPilot and PX4 (available at \cite{ardupilotHinj} and \cite{px4Hinj}). \code{libhinj} also provides a C interface so that developers can instrument other UAV firmware.

\subsubsection{Scheduler}
The scheduler is responsible for determining if a sensor instance should be failed and for recording mode transitions. The scheduler uses RPCs to communicate with \code{libhinj}. Here, we discuss how the scheduler implements Algorithm \ref{alg:Avis}. 

The scheduler records the fault injection scenarios it has already explored to prevent redundant exploration. The fault scheduler represents a fault injection scenario as a set of tuples $(\text{Timestamp}, \text{Fault})$, where the fault component describes the injected fault (e.g. sensor and instance) and the timestamp is the simulation time when the fault was injected. We store each scenario in a hash-set. The scheduler simulates a scenario if it does not already appear in the hash-set.

The scheduler uses algorithm \ref{alg:Avis} to determine the next fault scenario. When it is time to insert a failure, the scheduler responds to the RPC from \code{libhinj} indicating to fail the sensor read operation.

\subsection{Invariant Monitor}

At the end of each step, Avis's invariant monitor checks that the vehicle is operating correctly. Besides the UAV's mode reported by \code{libhinj}'s APIs, the invariant monitor also requires the vehicle's physical state, i.e., the position and the acceleration, for detecting invariant violation. The physical state of the vehicle is reported from the vehicle's Gazebo plugin over a Unix socket. We apply equation \ref{invariant:liveliness} to detect when liveliness is violated. Safety violations are reported using a simple crash detector.

\begin{table}[t]
    \centering
    \begin{tabular}{c|c|c|c|c}
         \hline
         \textbf{Features} & \textbf{Avis} & \textbf{Strat. BFI} & \textbf{BFI} & \textbf{Rnd}   \\
         \hline
         Targets operating mode transitions & \checkmark & \xmark & \xmark   &  \xmark \\
         Prior bugs inform injection sites  & \checkmark & \checkmark & \checkmark  & \xmark \\
         Search dissimilar scenarios first & \checkmark & \checkmark & \xmark & \checkmark \\
         \hline
    \end{tabular}
    \caption{Distinguishing features of Avis versus competing fault-injection approaches.}
    \label{tab:competingapproaches}
\end{table}

\section{Evaluation}
\label{sec:evaluation}

%

We evaluated Avis using two popular UAV platforms: ArduPilot's ArduCopter-3.6.9~\cite{ArduPilot} and PX4-1.9.0~\cite{px4}. We selected these systems because they are popular, sophisticated, and open-source~\cite{antwork,canberra}. We used the 3DR Iris quadcopter \cite{3driris} as the UAV in all experiments.  We selected the Iris because quadcopters are the most common body type used for UAVs and both ArduPilot and PX4 have robust support for the Iris. We conducted all experiments on a server equipped with CentOS 7.3, 8 GB of memory, and a quad-core Intel Xeon running at 2.66 GHz.

As shown in Table~\ref{tab:competingapproaches}, we compared 
Avis to three competing approaches. 
Random fault injection (Rnd) chose fault injection sites 
from all sensor readings with equal probability.  It also
chose failure scenarios for simulation randomly.
We implemented Bayesian Fault Injection (BFI), a state-of-the-art approach 
for injecting sensor faults in autonomous cars~\cite{bfi}. 
This approach used machine learning to predict which injection 
sites were most likely to trigger unsafe conditions.
We implemented BFI using depth-first search 
to explore injection scenarios.  However, BFI does not require depth-first search.
We also implemented an improved version of BFI called
Stratified BFI that uses SABRE to explore injection candidates using BFI's algorithm.  While Stratified BFI improved upon the state of the art, 
it missed a key feature of Avis.  Specifically, it did not exhaustively target the 
critical periods where UAV transitioned between operating 
modes.

We ran each approach for 2 hours per workload (see~\ref{sec:implementation}). 
First, we studied unsafe conditions uncovered by Avis, looking for 
previously unknown sensor bugs.  We also studied the unsafe 
conditions found by competing approaches to see if they revealed
the same sensor bugs.  This analysis shows the efficacy of Avis.
We compared the number of unsafe conditions found by each approach, 
a measure that reveals the efficiency (i.e., unsafe conditions per unit time).
We also re-inserted known bugs into the code base, ran each 
approach and looked for unsafe conditions caused by the known bugs.
Our evaluation also considered slowdown caused by Avis.


\begin{table*}[t!]
  \caption{Unknown bugs found by Avis.}
  \label{tab:detected-bugs}
  \centering
  \begin{tabular}{c|c|c|c|c|c|c}
    \hline
    \textbf{Report \#}
    & \textbf{Firmware} 
      & \textbf{Symptom}
        & \textbf{Sensor Failure} 
          & \textbf{Failure Starting Moment}
            & \textbf{Avis}
              & \textbf{Stratified BFI}
            \\
    \hline
    APM-16020
    & ArduPilot
    & Fly Away
      & GPS
        & \code{Takeoff $\rightarrow$ Autopilot}
          & \checkmark
            & \xmark
      \\
    APM-16021
    & ArduPilot
    & Crash
      & Accelerometer 
        & \code{Takeoff $\rightarrow$ Waypoint 1}
          & \checkmark
            & \checkmark
        \\
    APM-16027
    & ArduPilot
    & Fly Away
      & Barometer
        & \code{Pre-Flight $\rightarrow$ Takeoff} 
          & \checkmark
            & \xmark
        \\
    APM-16967
    & ArduPilot
      & Crash
        & Compass
          & \code{Waypoint 1 $\rightarrow$ Waypoint 2}
            & \checkmark
              & \checkmark
         \\
    APM-16682
    & ArduPilot
    & Crash
      & Accelerometer 
        & \code{Return To Launch $\rightarrow$ Land}
          & \checkmark
          & \xmark
        \\
    APM-16953
    & ArduPilot
    & Crash
      & Gyroscope
      & \code{Return to Launch $\rightarrow$ Land}
        & \checkmark
          & \xmark
    \\
    PX4-17046
    & PX4
    & Fly Away
      & Gyroscope
        & \code{Waypoint 3 $\rightarrow$ Return To Launch} 
            & \checkmark
              & \checkmark
     \\
    PX4-17057
    & PX4
    & Crash
      & Gyroscope
        & \code{Pre-Flight $\rightarrow$ Takeoff}
          & \checkmark
            & \checkmark
        \\
    PX4-17192
    & PX4 
    & Takeoff Failure 
      & Compass
        & \code{Pre-Flight $\rightarrow$ Takeoff}
          & \checkmark
          & \xmark
        \\
    PX4-17181
    & PX4 
    & Takeoff Failure
      & Barometer
        & \code{Pre-Flight $\rightarrow$ Takeoff}
          & \checkmark
            & \xmark
        \\
    \hline
  \end{tabular}
\end{table*}

\subsection{Detecting Unknown Bugs}


Table~\ref{tab:detected-bugs} lists the bugs detected by Avis. For each bug,
the table also shows the affected firmware (\textbf{Firmware}), the symptom of the bug (\textbf{Symptom}), the injected sensor failure (\textbf{Sensor Failure}), and the starting time of the fault (\textbf{Failure Starting Moment}).

In total, Avis discovered 10 previously unknown bugs: 6 affected ArduPilot and 4 affected PX4. These bugs were serious -- 2 that affected ArduPilot resulted in a vehicle crash and 3 made the UAV ignore user commands and fly away. A PX4 bug caused a crash and another caused a fly-away. The system logs showing unsafe behavior are available at \cite{avissource} in the \code{logs} directory.

The unsafe conditions that Avis found revealed sensor bugs triggered 
by GPS, accelerometer, barometer, compass and gyroscope failures.  
Manifestations of the newly found sensor bugs were also sensitive 
to timing conditions, a factor that explained why competing approaches
were unable to find them.  
Avis reported no false positives. However, each bug can manifest
multiple unsafe conditions.

\vspace{0.1in}
\noindent {\bf Case Study APM-16682:}
\label{subsec:case-study-1}
UAVs use fail-safe mechanisms to survive sensor failures, but 
sometimes simply triggering a fail-safe can yield unsafe conditions.  
Instead, the firmware should check flight conditions to ensure
fail-safe tasks can be supported. Recall Figure~\ref{fig:draw-bugexample},
an IMU fault during the landing mode triggered a fail-safe that
eventually caused a crash.  None of the competing approaches 
captured unsafe conditions caused by this sensor bug within a 2-hour run.  
The landing sequence represented less than 4 seconds of 
the 80 second scene (i.e. $\leq$ 5\%). Random fault injection 
must run for nearly 10 hours to achieve a 98\% certainty of capturing
a manifestation of the bug. BFI also failed to uncover this scenario, 
because the model learned by BFI did not include training data where 
unsafe conditions arose during landing.  
In contrast, AVIS uncovered this scenario in an hour.

\begin{figure}[b]
    \centering
    \includegraphics[width=\columnwidth]{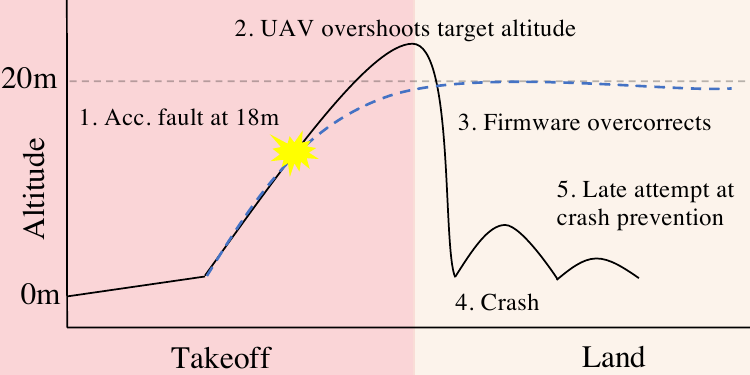}
    \caption{Sequence of events in APM-16021. 
    The black line shows the altitude of the UAV under fault injection. The blue line shows the altitude of the UAV during the golden run.}
    \label{fig:apm-16021}
\end{figure}

\vspace{0.1in}
\noindent {\bf Case Study APM-16021} 
\label{subsec:case-study-3}
Figure \ref{fig:apm-16021} shows 
APM-16021,
a new bug Avis found. The workload commanded the UAV to ascend to a target altitude of 20m. Before the UAV reached 20m, Avis injected an accelerometer fault (1). This caused the UAV to overshoot the target altitude (2). The firmware responded by landing (3). The firmware's state model incorrectly predicted a high altitude, causing it to allow the UAV to crash (4). The firmware made a final attempt to prevent the crash that had already occurred and unsafely actuated on the ground (5). Without any fault injection, the UAV's mode changed from takeoff to guided after it ascends to 20m. Avis detected this mode transition and injected faults around this time. Because our workload held the altitude constant inside the guided mode, an IMU fault at this time did not cause a crash. After several unsuccessful fault injections, Avis injected a fault at 18m and triggered the bug.

\begin{figure}[h]
    \centering
    \includegraphics[width=\columnwidth]{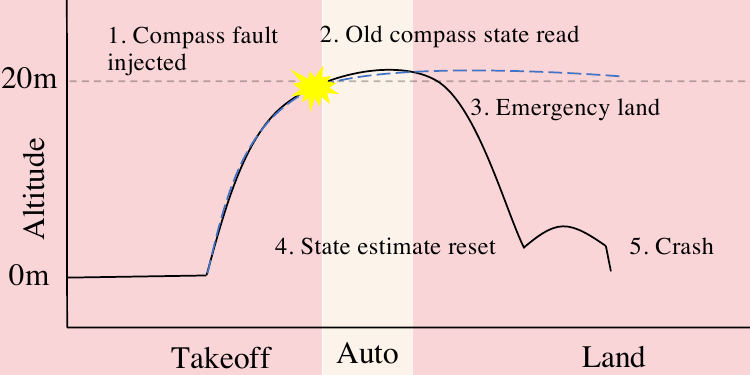}
    \caption{Sequence of events in
    APM-16967. 
    The black line shows the altitude of the UAV under fault injection. The blue line shows the altitude of the UAV during the golden run.}
    \label{fig:fnd-7}
\end{figure}

\vspace{0.1in}
\noindent {\bf Case Study APM-16967:}
\label{subsec:case-study-4}
Figure \ref{fig:fnd-7} shows an unsafe condition found by both Avis and Stratified BFI. This bug is triggered if a compass fails anytime between waypoints. Avis triggered this bug by injecting a compass failure after the UAV reached waypoint 1 (1). Then, the vehicle turned to fly towards its second waypoint. As the UAV turned, the firmware continued to use old compass readings (2). This caused the firmware to lose its heading estimate. The land fail-safe activated (3). The firmware reset its state estimate near the end of the landing procedure (4) which caused a crash (5). Stratified BFI is able to trigger this bug because its training data contains examples of compass failures in the body of the auto mode, but vanilla BFI does not reach this state.

\subsection{Comparison with Alternative Approaches}

\begin{table}[t]
    \centering
    \begin{tabular}{c|c|c|c}
         \hline
         \textbf{Approach} & \textbf{ArduPilot Unsafe \#} & \textbf{PX4 Unsafe \#} & \textbf{Total \#}   \\
         \hline
         Avis & 104 & 61 & 165 \\
         Strat. BFI & 61 & 9 & 70 \\
         BFI & 1 & 1 & 2 \\
         Random & 2 & 3 & 5 \\
         \hline
    \end{tabular}
    \caption{Unsafe scenarios identified by each approach.}
    \label{tab:mainresults}
\end{table}

Table \ref{tab:mainresults} reports the number of unsafe conditions 
identified by each approach.  Recall, each approach was run for 
two hours.  Avis found  more than 2.4X more unsafe conditions than stratified BFI, 
an improved implementation of the current state-of-the-art. 
Avis found 82X more conditions than BFI using standard depth-first search. 

\begin{table}[t]
    \centering
    \begin{tabular}{c|c|c|c|c}
         \hline
         \textbf{Approach} & \textbf{Takeoff \#} & \textbf{Manual \#} & \textbf{Waypoint \#} & \textbf{Land \#} \\
         \hline
         Avis & 60 & 37 & 44 & 24 \\
         Stratified BFI & 4 & 32 & 35 & 1 \\
         BFI & 1 & 1 & 0 & 0 \\
         Random & 0 & 2 & 3 & 0 \\
         \hline
    \end{tabular}
    \caption{Number of unsafe scenarios identified by each approach in each mode.}
    \label{tab:unsafe-scenarios-by-mode}
\end{table}

BFI did not uncover many unsafe conditions for two reasons.  
First, depth-first search inefficiently checked fault injection scenarios
that were effectively redundant.  The 3D Iris sampled sensors $10^3$ to $10^4$ 
times per second. In our experiments, BFI's model took $\sim10$ seconds to label an injection scenario. 
BFI was unable to explore even a single second of data within its 2 hour budget.
Stratified BFI addresses this 
problem by using SABRE, Avis's injection schedule.  However, Stratified BFI failed to correctly predict the behavior of sensor failures during modes that were not executed long in the workload.  Table \ref{tab:unsafe-scenarios-by-mode} shows a breakdown of unsafe scenarios found in each mode
by each approach.

\subsection{Detecting Existing Bugs}
In order to approximate Avis' false negative rate, we evaluated Avis using bugs that were 
previously reported on Github.  We used 
5 sensor bugs that (1) did not require special
environmental conditions, e.g., heavy winds, 
(2) applied to the Iris quadcopter and (3) had
serious symptoms.  We reinserted these bugs into
the code base and used Avis to find unsafe conditions.
As shown in Table~\ref{tab:existing-bugs-found},
Avis found unsafe conditions triggered by all 5 bugs. 
Stratified BFI found 2.  BFI and random found none.
Further, Avis triggered the bugs quickly, using at
most 21 simulations.  Stratified BFI, using the 
SABRE search algorithm, also discovered bugs 
quickly when it was effective.

\begin{table}[t]
    \centering
    \begin{tabular}{c|c|c|c|c}
        \hline
        
        \textbf{Bug ID} & \multicolumn{2}{c|}{\bf Avis} & \multicolumn{2}{c}{\bf Strat. BFI} \\
                        \cline{2-5}
                        & Found & {Simulations} & Found & {Simulations} \\
        \hline
        APM-4455 & \checkmark & 10 & \xmark & N/A \\
        APM-4679 & \checkmark & 21 & \checkmark & 3 \\
        APM-5428 & \checkmark & 5  & \xmark & N/A \\
        APM-9349 & \checkmark & 4  & \checkmark & 5 \\
        PX4-13291& \checkmark & 18 & \xmark & N/A \\
        \hline
    \end{tabular}
    \caption{Existing bugs triggered by Avis.}
    \label{tab:existing-bugs-found}
\end{table}

Table~\ref{tab:existing-bugs-found} shows that 
Stratified BFI does not identify bugs that require multiple failures, like PX4-13291. PX4-13291 reports that a fly-away occurs when the UAV's battery drops to an unsafe level without local position. Avis triggers this bug by injecting a GPS fault. This causes the UAV to lose its local position estimate. Then, Avis injects a battery sensor failure. This causes the firmware to trigger the battery fail-safe. At this point, Avis has triggered the bug. Stratified BFI does not uncover this scenario because the UAV handles the GPS or battery failure, but not both together. Having not seen the effects of joint failures in the training data, the model is unable to predict this outcome.

%


%


\section{Related Work}
\label{sec:related-work}

Our work is related to in-situ model checking, cyber-physical attacks and mitigation, sensor fault detection, and empirical UAV bug studies.

\paragraph{In-Situ Model Checking} In-situ model checkers have been successfully used to check many real systems such as network protocol implementations \cite{cmc} and file systems \cite{explode,fisc}. More recently, this technique has been applied to distributed computing \cite{dir,samc,modist,crystalball,fate-destini}. However, existing in-situ model checkers are not effective in triggering sensor bugs in UAVs because they do not consider the behavior of the vehicle. As the first in-situ model checker designed for UAVs, Avis injects sensor faults at mode boundaries to be both effective and efficient at triggering sensor bugs.

\paragraph{Cyber-Physical Attacks and Mitigation}
UAVs have become a hot topic of security research in recent years. Sensors have been shown to be a major attack vector for UAVs \cite{rocking-drones,walnut,CYT}. These attacks work by disturbing the UAV's sensors to cause their models to diverge from their physical state and actuate according to the attacker's desires. RVFuzzer \cite{rvfuzzer} shows a method to measure the similarity between mission executions by using state deviation, an idea that we refined in our own fly-away detector. A large body of work demonstrates how to use control semantics to detect or mitigate attacks against vehicle \cite{crosslayer-retrofitting,control-invariant,safd}. We rely on similar principles to design our pruning policies.

\paragraph{Detecting Sensor Faults}
Prior research shows how to detect byzantine sensor faults. Control semantics are used for this in \cite{safd}. \cite{nnsensorfaults} and \cite{sensorfault-tolt} leverage neural networks to detect or mitigate sensor faults. Recently, \cite{currentSense} shows how to use physical measurements to detect sensor faults. Prior work shows how vehicles can survive our fault model \cite{fault-det-rec}. Our work demonstrates how to detect when a UAV fails to correctly handle sensor failures.

\paragraph{Empirical RV Bug Studies}
Both \cite{space-fault-types} and \cite{crashing} look at bug reproducibility. They find that UAV bugs are reproducible. We provide similar data for sensor bugs specifically and show they are reproducible. \cite{robocup} shows that sensor bugs afflict participants in the robotic soccer league.

\section{Conclusion}
\label{sec:conclusion}
Unmanned aerial vehicles rely on sensors to model their 
physical states and must contend with sensor failures. 
Our empirical bug study on ArduPilot and PX4, two popular
open-source UAV control firmware, showed severe consequences 
for mishandling sensor failures, a.k.a, sensor bugs.  
We presented AVIS: an in-situ model checking approach for UAVs. 
Even though UAVs access sensors frequently and many sensor bugs
manifest only if failures occur within narrow timing windows, 
we used AVIS to find 10 previously unknown sensor bugs of which 2 have been
reproduced by firmware developers.  Avis used
modern firmware support for operating modes to inject 
sensor failures at critical points during flight execution.
Avis provides a missing tool for software developers, enabling
a preemptive approach to diagnose sensor bugs and analyze their
root causes.  We hope our work improves reliability for this
emerging technology and unlocks new UAV applications.
In addition, we hope our work can draw more attention to 
UAV reliability in the community.



\bibliographystyle{plain}
\bibliography{sample-base,cstewart.bib}


\end{document}